\documentclass[aps,prb,twocolumn,amsmath,amssymb,superscriptaddress,floatfix,showpacs]{revtex4}
\usepackage[dvips]{graphicx}
\usepackage[T1]{fontenc}
\usepackage[latin1]{inputenc}
\usepackage[active]{srcltx}
 
\begin{document}
\title{Classical correlations of defects in lattices with geometrical 
frustration in the motion of a particle.}

\author{Frank Pollmann}
\affiliation{Max--Planck--Insitut f{\"u}r Physik komplexer Systeme, N{\"o}thnitzer Str. 38, 01187 Dresden, Germany}
\author{Joseph J. Betouras}
\affiliation{Max--Planck--Insitut f{\"u}r Physik komplexer Systeme, N{\"o}thnitzer Str. 38, 01187 Dresden, Germany}
\affiliation{Instituut Lorentz, Universiteit Leiden, Niels Bohrweg 2, 2333 CA Leiden, The Netherlands}
\author{Erich Runge}
\affiliation{Technische Universit\"at Ilmenau, Institut f\"ur Physik, 98684 Ilmenau, Germany}

\date{\today}
\begin{abstract}
We map certain highly correlated electron systems on lattices with geometrical frustration in 
the motion of added particles or holes to the spatial defect-defect correlations of dimer 
models in different geometries. These models are studied analytically and 
numerically. We consider different coverings for four different lattices: square, honeycomb, 
triangular, and diamond. In the case of hard-core dimer covering, 
we verify the existed results for 
the square and triangular lattice and obtain new ones for the honeycomb and the diamond lattices 
while in the case of loop covering we obtain new numerical results for all the lattices and use
the existing analytical Liouville field theory for the case of square lattice.
The results show power-law correlations for the square and honeycomb lattice, 
while exponential decay with distance is found for the triangular and exponential decay with 
the inverse distance on the diamond lattice. We relate this fact with the lack of bipartiteness 
of the triangular lattice and in the latter case with the three-dimensionality of the diamond. 
The connection of our findings to the problem of fractionalized charge in such lattices is 
pointed out. 
 \end{abstract}
\maketitle

\section{Introduction}
There has been enormous interest in the properties of quantum frustrated magnets, with recent 
studies showing fractionalization of quantum numbers (see Refs. [\onlinecite{diep, balents}], 
and citations therein). More recently, the attention has shifted to charge degrees of freedom 
on lattices with frustrated geometries where the number of classical ground state configurations 
increases exponentially with the number of sites. It has been noticed that hopping of fermions 
from site to site in frustrated lattices can lead, under certain conditions, to 
fractionalization of the fermionic charge in a rather natural way. \cite{pfulde1,pfulde2}

The motion of spinless fermions (or hardcore bosons) on a lattice can be mapped to an Ising 
spin problem or to a dimer configuration for any given lattice geometry and filling fraction 
as we will see in the next paragraph. The addition of a fermion (or boson) leads to a flip of 
an Ising spin or, equivalently, to a new dimer. The ground states of the considered systems 
fulfill a local constraint of having a certain number of particles on each unit cell or, 
equivalentely, a certain number of dimers at each site.

\begin{figure}
\begin{center}
\includegraphics[width=65mm]{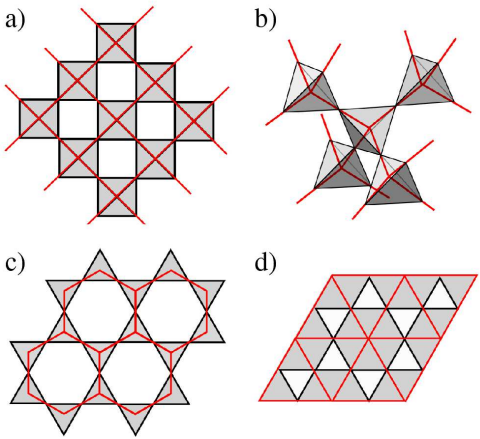}
\end{center}
\caption{(color online) Mapping of lattices which show frustration (black) to dual lattices 
(red). (a) Checkerboard lattice with nearest-neighbor repulsion $\rightarrow$ square lattice, 
(c) pyrochlore lattice with nearest-neighbor repulsion $\rightarrow$ diamond lattice 
(c) kagome with nearest-neighbor repulsion $\rightarrow$ honeycomb lattice and 
(d) kagome with repulsion interaction on hexagons only $\rightarrow$ triangular lattice. 
The constraint of having a certain number of particles on each unit cell translates into the 
constraint of having a fixed number of dimers touching each site of the dual lattice, 
leading to a hard-core dimer or loop dimer covering.
\label{cap:Mapping}}
\end{figure}

For a specific implementation of local constraints in the Hamiltonian language for fermions, 
let us consider the model Hamiltonian
\begin{equation}
H=-t\sum_{\langle i\  j\rangle}\left(c_{i}^{\dag}c_{j}^{\vphantom{\dag}}+\mbox{H.c.}\right)+
V\sum_{\langle i\  j\rangle}n_{i}n_{j}\ .\label{eq:hamil}\end{equation}
Here $\langle i\  j\rangle$ denotes the sum over nearest neighbors $i$ and $j$. The operators 
$c_{i}^{\vphantom{\dag}}(c_{i}^{\dag})$ annihilate (create) a particle on site $i$, and 
$n_{i}=c_{i}^{\dag}c_{i}^{\vphantom{\dag}}$. For a moment, let us set the hopping integral $t$ 
to zero. We are interested in lattices for which the ground state of the Hamiltonian 
(\ref{eq:hamil}) with repulsive nearest-neighbor interaction term $V$ only has at certain 
fillings a macroscopic degeneracy, which increases exponentially with the system size. In other 
words, the system has a finite $T=0$ entropy (for possible technical applications, see 
Ref.~[\onlinecite{zhitomirsky}]). Examples of such lattice are the pyrochlore lattice and the 
criss-crossed checkerboard lattice (two dimensional pyrochlore), shown as 
Fig.~\ref{cap:Mapping}~(a) and (b). At half filling, all classical ground states fulfill 
the so-called tetrahedron rule of having exactly two particles on each tetrahedron 
(criss-crossed square) \cite{anderson}. The dimer coverings of the dual lattice is constructed 
as usual by connecting all centers of the tetrahedra and drawing a dimer whenever the traversed 
site of the original lattice is occupied. The tetrahedron rule translates into the constraint 
of having exactly two dimers at each site of the dual lattice. Similarly, for the kagome 
lattice with nearest-neighbor repulsion the dual lattice is the honeycomb, see 
Fig.~\ref{cap:Mapping} (c). If the repulsive interaction on the kagome lattice is restricted 
to hexagons only, the dual lattice is a triangular lattice, see Fig.~\ref{cap:Mapping} (d).

In the long-run, one wants to understand the behavior of mobile, fractionally charged, 
excitations resulting from weak particle or hole doping of such systems.

As a first step, we present the study of classical defect-defect correlations. A defect here 
means a cell where the rule of fixed number of particles (or dimers) is violated (Fig. 2). 
The study includes bipartite and non-bipartite two-dimensional lattices (checkerboard, 
honeycomb and triangular lattices) as well as a three-dimensional diamond lattice, with two 
different fillings corresponding to (i) the loop dimer and the (ii)hard-core dimer covering. 
Wherever possible, we obtain analytical results and compare them to numerical simulations. 
These  classical correlations provide informations about correlations near the Rokshar-Kivelson 
point (RK point) of the quantum Hamiltonian because at the RK point the quantum mechanical 
ground state is given by an equal weighted superposition of all configurations \cite{RK}. 
Furthermore, in our case, one can expect classical correlations to hold more generally since 
the low-energy excitations can be described equivalently by hardcore bosons or fermions 
\cite{frank}.

\begin{figure}
\begin{center}\begin{tabular}{cc}
(a)\includegraphics[%
  width=30mm]{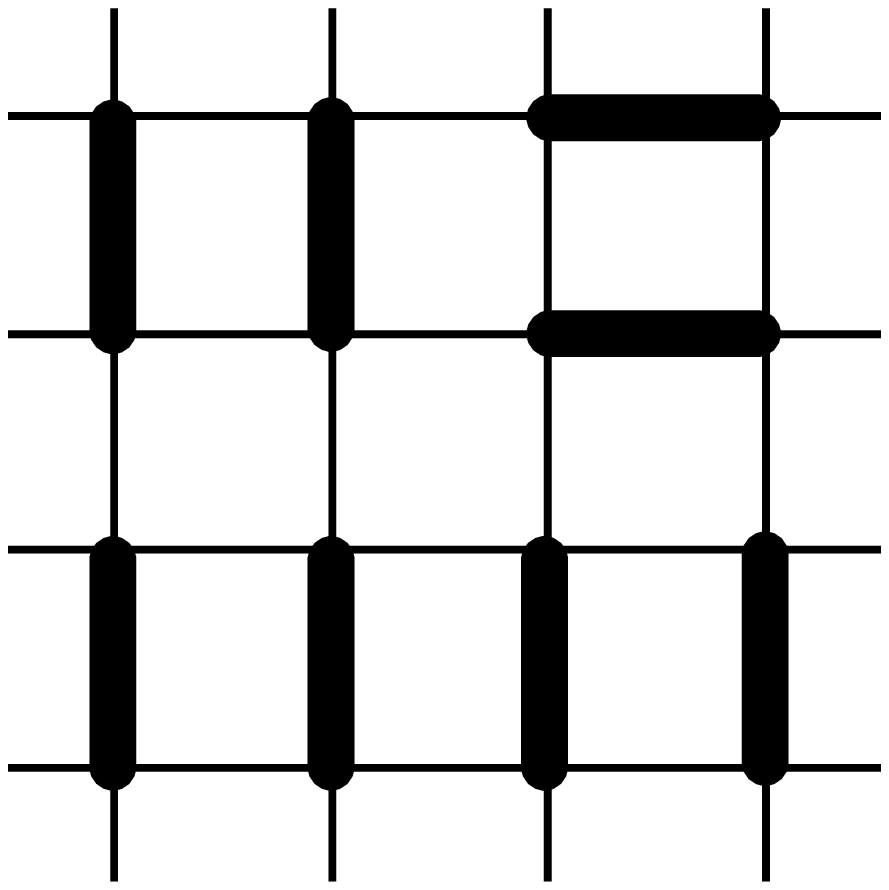}&
(b)\includegraphics[%
  width=30mm]{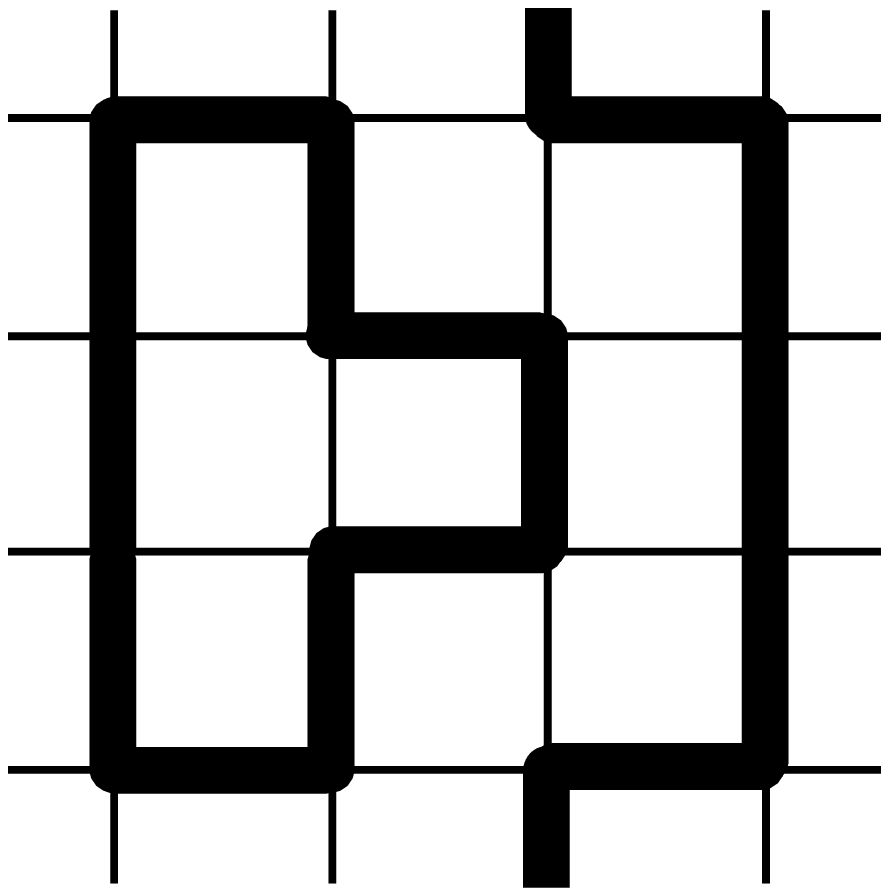}
\tabularnewline
(c)\includegraphics[%
  width=30mm]{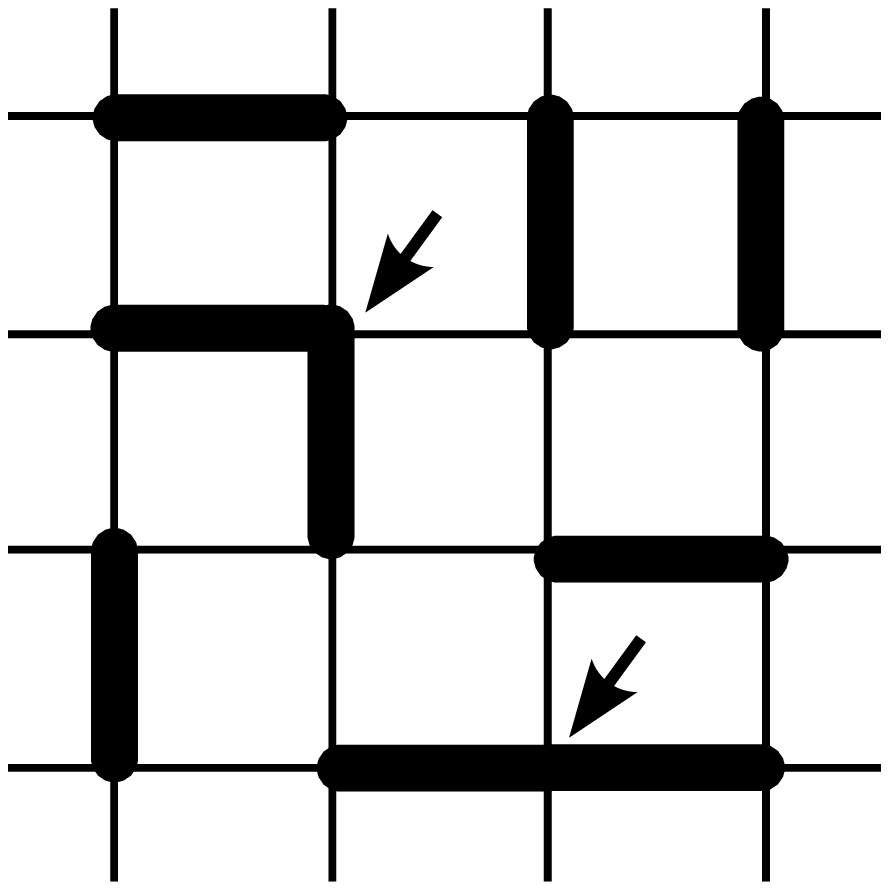}&
(d)\includegraphics[%
  width=30mm]{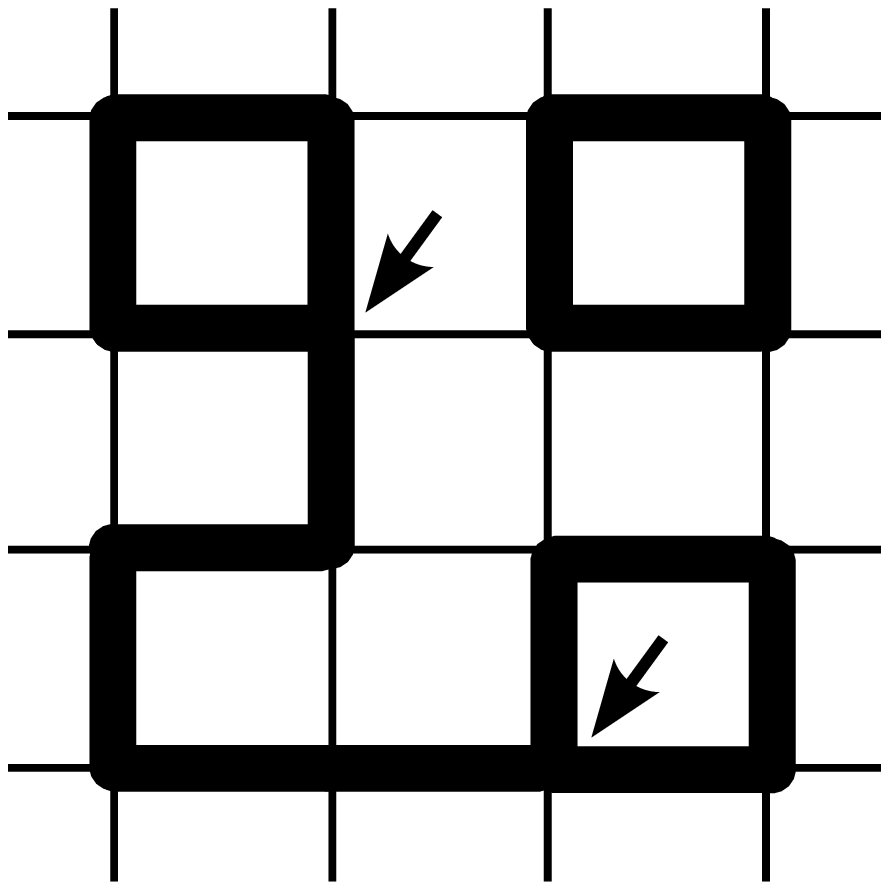}
\end{tabular}\end{center}
\caption{Hard-core dimer (a) and loop dimer (b) covering on the square lattice. Figures (c)
and (d) show two defects in the different coverings at postions marked by
arrows.
\label{fig2}}
\end{figure}

Historically, studies of monomer and dimer models started very early \cite{fowler}. At zero 
doping, the Pfaffian method introduced by Kasteleyn allows to obtain classical correlations 
analytically \cite{kasteleyn}. For the square lattice with hard-core dimer covering, 
the correlation of a pair of monomers has been obtained \cite{fisher,hartwig}. 

The main findings of the present work are: For the square and the honeycomb lattice, 
correlations as function of distance show a power-law behavior, with an exponent 1/2 for the 
hard-core dimer covering on the square lattice and 1/3 for a loop dimer covering. 
The honeycomb lattice has an exponent of 1/2 for both coverings. 
An exponential decay with 
distance is found for the triangular lattice and the diamond lattice 
shows exponential decay with the \textit{inverse} distance.

In the next section, we present the results and the methods in detail. We start with the square 
lattice where, for completeness, we summarize the Liouville theory of the fully packed loop 
model with two colors \cite{kondev}, which provides the exponent for the loop covering 
analytically and we subsequently show the numerical results. We then present the relevant 
results for the honeycomb, the triangular and the diamond lattices and finally, we conclude.

\section{Results for different lattices}
\subsection*{A. Square lattice}

\textit{Square lattice}. As described above, the half-filled checkerboard lattice with 
nearest-neighbor repulsion is mapped to a loop dimer model on the square lattice, meaning that 
each site is connected to exactly two dimers. This can then be interpreted as covering in the 
fully packed loop model with two colors, assigning different colors to occupied and unoccupied 
bonds. Using the machinery developed for this model, \cite{kondev} we can read off the 
appropriate exponent of the defect-defect correlation function: In order to obtain the 
Liouville theory of the model, the first step is to map the oriented loops to an interface 
model, where they can be interpreted as contour lines. The microscopic heights are defined at 
the center of the lattice plaquettes and each bond is in one of four possible states 
($\vec{A}$, $\vec{B}$, $\vec{C}$, $\vec{D}$). We write the partition function as 
$Z= \sum_G n_b^{N_b} n_g^{N_g}$ where $N_b$ and $N_g$ are the numbers of the loops of the two 
colors and $0 \leq n_b,\; n_g \leq 2$ are the fugacities accordingly. Assigning weights 
$\exp(\pm i \pi e_b)$ or $\exp(\pm i \pi e_g)$ to the different orientations, with 
$n_b= 2 \cos(\pi e_b)$ and $n_g= 2 \cos(\pi e_g)$, allows a local definition of the statistical 
weight of oriented loop, and in turn the weights leads to a local field theory.\cite{kondev}

Our problem at hand (which can be equivalently mapped to a six vertex model), corresponds to 
fugacities $n_b$=$n_g$=$1$. Therefore $e_b$=$e_g$=$1/3$. Adopting the same convention as in 
Ref. [\onlinecite{kasteleyn}]  [$\vec{A} =(-1,+1,+1)$, $\vec{B} =(+1,+1,-1)$, 
$\vec{C} =(-1,-1,-1)$, $\vec{D} =(+1,-1,+1)$], the topological charge ($\pm \vec{m}$) 
is $\vec{m} = \vec{C}-\vec{A} = (0,-2,-2)$ and $2 x_{1,1}$ is the dimension of an operator 
with total charge $(\vec{e}_0,\vec{m}_{1,1})$, 
where $\vec{e}_0=-\pi/2(e_g+e_b,0,e_g-e_b)= \pi(1/3,0,0)$. 
The effective field theory for the coarse-grained heights is given by the action

\begin{eqnarray}
S&=&S_E+S_B+S_L \nonumber \\
S_E&=& \frac{1}{2} \int d^2x K_{\alpha,\beta} \partial h^{\alpha} \partial h^{\beta} \nonumber \\
S_B&=& \frac{i}{4 \pi} \int d^2 x (\vec{e}_0 \cdot \vec{h}) {\cal R} \nonumber  \\
S_L&=& \int d^2 x \sum_{\vec{e} \in R_{\omega}^{*}} 
{\tilde{\omega}}_{\vec{e}}\ exp(i \vec{e} \cdot \vec{h}(x)).
\nonumber 
\end{eqnarray}
Here, $S_E$ is the elastic term (entropy of the oriented loops) with the tensor $\cal K$ of 
elastic constants $K_{11}=K_{33}$ and $K_{12}=K_{23}=0$ and 
$\vec{h}(\vec{x})$ is the height field (coarse grained field of the integer-valued height h). 
$S_B$ is the boundary term with the scalar curvature  ${\cal R}$, which vanishes everywhere 
except at the boundary. This term inserts the vertex operator at far ends and supplies winding 
loops. Finally, $S_L$ is the Liouville term. 
The Liouville potential is the coarse grained version of the local weight assigned to an 
oriented loop configuration 
$\prod_{\vec{x}} \lambda(\vec{x})$: $\omega(\vec{x})=-\ln(\lambda(\vec{x}))$ is defined in 
height space, which is the bcc lattice $R_{\omega}$. The Fourier-transformed values of 
$\omega(\vec{x}) = \sum_{\vec{e} \in R_{\omega}^{*}}{\tilde{\omega}}_{\vec{e}}\ \exp(i \vec{e} 
\cdot \vec{h}(x))$ in the reciprocal lattice $R_{\omega}^{*}$ (a fcc lattice) enter $S_L$. 

The scaling dimension of a general operator with both electric and 
magnetic charge is given by:
\begin{equation}
x(\vec{e},\vec{m})= \frac{1}{4 \pi} \vec{e} \cdot [ {\cal K}^{-1} \cdot (\vec{e}-2 \vec{e}_0) + 
\frac{1}{4 \pi} \vec{m} \cdot [ {\cal K} \cdot \vec{m} ]
\end{equation}
where in our case $K_{11}=\frac{\pi}{8} (2-e_b-e_g)= \frac{\pi}{6}$, $K_{13}=\frac{\pi}{8} 
(e_b-e_g)=0$
and $K_{22} = \frac{\pi}{2} \frac{(1-e_b)(1-e_g)}{2-e_b-e_g}=\frac{\pi}{6}$.
Therefore the critical exponent of interest is simply:
\begin{eqnarray}
2 x_{1,1}&=& \frac{1}{4} [ (1-e_b)+ (1-e_g)] + \frac{(1-e_b)(1-e_g)}{2-e_b-e_g} \nonumber \\
& & - [\frac{e_b^2}{1-e_b}+ \frac{e_g^2}{1-e_g}] = \frac{1}{3}. \label{onethird}
\end{eqnarray}
The result (\ref{onethird}) will be next verified numerically by Monte Carlo simulations. 

%
The dimer models on dual lattices are numerically preferable over the original lattices, 
because the constraints are included
in a more natural way. Even though dimer models have been intensively investigated in the 
study of spin models\cite{diep} not much is known about defect-defect correlations. 
The defect-defect correlation function, 
e.g. $C(\mathbf{0}, \mathbf{r})\sim \langle N_{\mathbf{0}}N_{\mathbf{r}}\rangle$ which gives 
the probability to find two defects at a certain distance $\mathbf{r}$. The defect-defect 
correlation is proportional to the restricted partition function 
$Z_{11}(\mathbf 0, \mathbf r)\sim |\mathbf{r}|^{-2x_{11}}$ of the two color fully packed 
loop model. It is counting the number of configurations with defects at 
$\mathbf 0$ and $\mathbf r$ connected by one string of each color.  
Thus $C(\mathbf{0}, \mathbf{r})$ is expected to become isotropic at large distances and to 
decay algebraically $Z(\mathbf{0}, \mathbf{r})\sim |\mathbf{r}|^{1/3}$.

In the following, we measure numerically the correlations along a coordinate axis 
[$\mathbf{r}=(x,0)$] and refer to it as $C(x)$. The classical two-point correlation functions 
at zero temperature are delivered as averages over all degenerate ground states. The number of 
degenerate ground states grows exponentially with the system size. At the start, an allowed 
configuration with fixed filling, i.e. a ground state of the undoped system with no violation 
of any local constraint is generated. Then we add a dimer onto an unoccupied random link. This 
leads to two defects on adjacent sites of the dual lattice, which subsequently propagate via 
local dimer moves through the system without creating any new defects. At each step the 
defect-defect distance $x$ is counted in a histogram $Z(x)$, which after normalization yields 
the correlation function $C(x)=Z(x)/Z_0$. The normalization is somewhat arbitrary, we chose 
it in such way that $C(1)=1$. The algorithm is terminated when the standard deviation of the 
measured quantity falls below a certain threshold. Results of simulations with different 
initial configurations are compared for verification. For the 2D lattices about 
$10^7-10^8$ samplings were necessary. For the 3D diamond lattice it took about 
$10^{10}$ steps until convergence. We also applied an alternative Monte Carlo algorithm with 
loop updates\cite{sandvik} which is known to be ergodic and unbiased but shows considerable 
slower convergence.

\begin{figure}
\includegraphics[width=80mm]{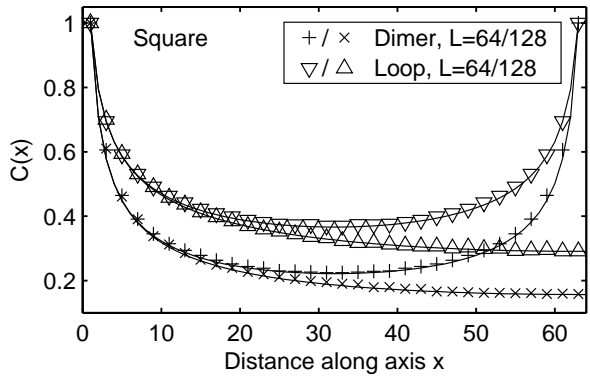}
\caption{Classical defect-defect correlation functions on a square lattice with hard-core dimer 
(upper panel) and loop dimer coverings (lower panel) along a coodinate axis.  Results are shown 
for lattices with $L=64$ and $L=128$. The numerical data is fitted by the exact asymptotic 
results. The periodic boundary conditions are taken into account by plotting the fit against a 
rescaled distance variable. \label{cap:square} }
\end{figure}

The algorithms have been applied to different lattice structures and filling factors 
(see Fig.~\ref{cap:Mapping}). To test the implementation of the algorithm, we first reproduced 
the known monomer two-point correlations on the hard-core dimer covering on the square and 
triangular 
lattice (see Fig.~\ref{cap:square} and Fig.~\ref{cap:triangular}) as well as the dipolar 
correlations in the undoped system on a square lattice.\cite{roderich1, isakov}

The square lattice is a bipartite lattice and the two defects are on different sublattices. 
We extracted the exponent from the numerical data by linear interpolation of $\log$-$\log$ 
plots and verified the results by finite size scaling $C(x/L)=L^{\gamma} (x/L)^{\gamma}c(x/L)$ 
with exponent $\gamma$ and system size $L$. Fig.~\ref{cap:square} compares the numerical data 
and the analytical results. For a rescaled distance variable $x^{'}=L\sin (\pi x/L)/\pi$ is 
used the numerical fit in order to account for the periodic boundary conditions. 

In the case of a hard-core dimer covering (quarter-filled checkerboard lattice) the exponent 
agrees with the results from Ref. [\onlinecite{roderich2}], i.e., $C(x)\sim 1/x^{1/2}$. The 
correlations in the case of a loop dimer covering (half-filled checkerboard) are very well 
fitted with the power law $1/x^{1/3}$, obtained analytically (see above). As expected, the 
correlations do not show any angle dependence at large distances. In both cases, the decay of 
the correlation function is algebraic which is expected for two-dimensional bipartite lattices: 
The two defects have long-range correlation and feel each others presence at all distances. 
The bipartiteness is also seen by the strictly zero correlations C(x) for distances which 
connect sites that reside in the same sublattice.

%

\subsection*{B. Honeycomb lattice}
\begin{figure}
\includegraphics[width=80mm]{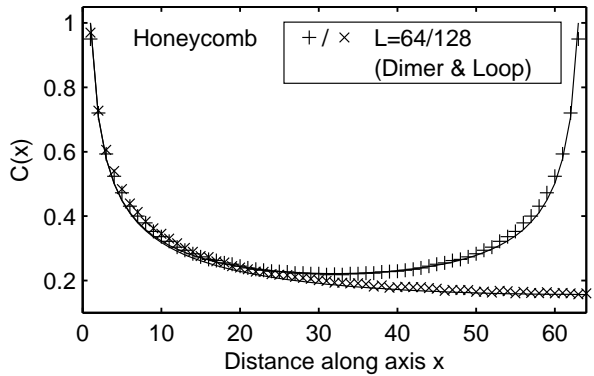}
\caption{Classical defect-defect correlation functions on a honeycomb lattice with hard-core 
dimer and loop dimer coverings along an axis of the bravais lattice.  Numerical results are 
shown for lattices with $L=64$ and $L=128$ together with to a power law $\sim x^{1/2}$. The 
periodic boundary conditions are taken into account by plotting the fit against a rescaled 
distance variable.
\label{cap:honeycomb} }
\end{figure}

The kagome lattice with nearest neighbor repulsion at one third and two third filling are 
mapped to the hard-core dimer and the loop dimer model on the honeycomb lattice, respectively. 
The numerically obtained correlations along the axes of the bravais lattice are both 
$C(x)\sim 1/x^{1/2}$ (see Fig.~\ref{cap:honeycomb}). Note that the two models, i.e., 
filling factors, are equivalent in the absence of defects. They can be identified by 
exchanging links which are occupied by a dimer and those which are not occupied. The faster 
algorithm described above did not show the equivalence of the two models. Therefore, we used 
the loop algorithm for this lattice. 
\subsection*{C. Triangular lattice}

\begin{figure}
\includegraphics[width=80mm]{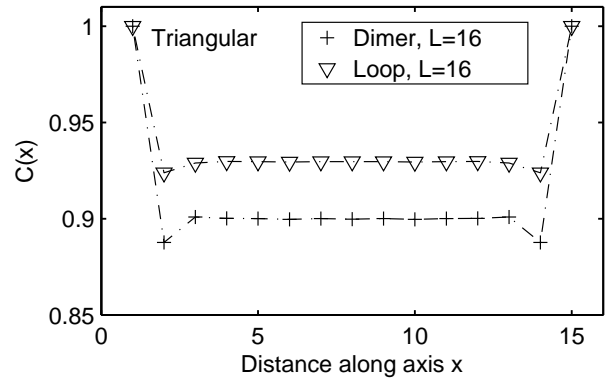}
\caption{Classical defect-defect correlation functions on a triangular lattice with hard-core 
dimer (upper panel) and loop dimer coverings (lower panel) along a coodinate axis.  
Results are shown for a lattice with $L=16$. \label{cap:triangular}}
\end{figure}

The kagome lattice with repulsion on the hexagons only is mapped at one sixth 
filling to a hard-core dimer model and at one third filling to a loop dimer model on the 
triangular lattice. The correlation function decays exponentially in both cases 
(see Fig.~\ref{cap:triangular}) with decay length of the order of one lattice spacing. 
The values obtained for the hard-core dimer model are in agreement with those obtained in Ref.  
[\onlinecite{roderich3}]. For distances further than a few lattice spacing $C(x)$ is constant 
within the noise ratio and tends to a finite value in the limit $x\rightarrow \infty$ 
\cite{roderich4}.  This implies that the free energy difference 
$\Delta F(\infty) \sim -T(\ln C(\infty)- \ln C(1))$ of two infinitely separated defects in 
the classical hard-core dimer and loop dimer model is finite and the two defects are deconfined.


%

\subsection*{D. Diamond lattice}
The mapping of particles with local constraints due to large repulsive interaction to hard-core 
dimer or loop loop models can be applied in three dimensional systems as well. For example, 
the pyrochlore lattice at one fourth and one half filling corresponds to the hard-core dimer 
and the loop dimer model on the diamond lattice, respectively.

\begin{figure}
\includegraphics[width=80mm]{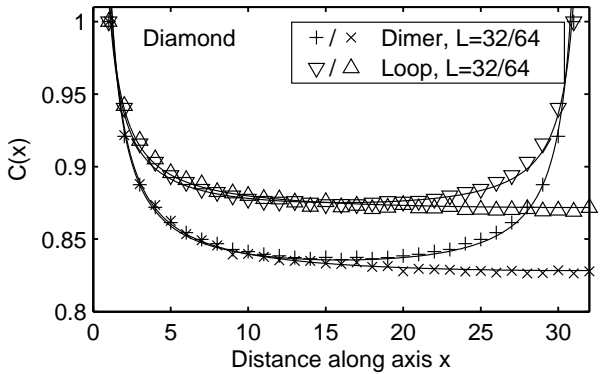}
\caption{Classical defect-defect correlation functions on a diamond lattice with hard-core 
dimer (upper panel) and loop coverings (lower panel) along a coodinate axis.  Results are 
shown for lattices with $L=32$ and $L=64$. The numerical data are fitted to exponential 
functions. The periodic boundary conditions are taken into account by plotting the fit against 
a rescaled distance variable.
\label{cap:diamond}}
\end{figure}

Monte Carlo simulations for two different lattice sizes are shown in Fig.~\ref{cap:diamond}. 
Much more samplings are needed to achieve a good signal to noise ratio, because the phase space 
is considerable larger than in the case of the 2D lattice. From a logarithmic plot, we found 
that the correlation functions decay approximately exponentially with respect to the inverse 
distance along the axes of the bravais lattice as $C(x)\sim \exp(1/4x)$ for the hard-core dimer 
covering and $C(x)\sim \exp(1/6x)$ for the loop covering. The reason that this bipartite 
lattice shows these short range correlations is because the monomers follow a 3D Coulomb law 
with potential ($V(r) \sim 1/r$) as opposed to the divergence at large distances $V(r) \propto 
\log r$ found in 2D \cite{huse}. The restricted partition function then decays exponentially 
with the distance in the case of the diamond lattice. A recent work on spin 1/2 Heisenberg 
antiferromagnet on the pyrochlore lattice \cite{hermele} reveals a fractionalized spin liquid 
with U(1) gauge structure where this $1/r$ potential acts between pairs of spinons and pairs of 
monopoles.


\section{Discussion}
%
In summary, we calculated the classical defect-defect correlations for different fillings and 
different lattices with geometry that causes frustration in the motion of particles. Recent 
studies on bipartite and non-bipartite three-dimensional lattices \cite{huse} lead to the 
conjecture that extended critical phases are realized only in bipartite lattices. This is 
supported by our results: For two-dimensional bipartite lattice/dual lattice pairs 
(checkerboard/square lattice and kagome/honeycomb lattice) the correlation functions decay with 
distance following a power law and tend to zero for $x\rightarrow \infty$. Otherwise 
("modified kagome with repulsive interaction on hexagons only/triangular lattice and three 
dimensional pyrochlore/diamond lattice") an exponential decay with distance or inverse distance, 
respectively, and a finite value for $x\rightarrow \infty$ is found.

\begin{table}
\begin{tabular}{|r|r|r|}
\hline
lattice&hard-core&loop\\
\hline
square&$x^{-1/2}$&$x^{-1/3}$\\
triangle&$exp(-x)$&$exp(-x)$\\
honeycombe&$x^{-1/2}$&$x^{-1/2}$\\
diamond&$exp(1/4x)$&$exp(1/6x)$\\
\hline
\end{tabular}
\caption{The summary of the defect-defect correlations we obtained for the four different 
lattices and the two distinct dimer coverings. In the case of the triangular lattice there is 
an exponential decay with decay distance of one lattice constant, tending to a finite value at
$x \longrightarrow \infty$.}
\end{table}

Where analytical values for power-law exponents were known in the literature, these could be 
confirmed. For the checkerboard lattice, i.e., two-dimensional pyrochlore lattice, we could 
relate the defect-defect correlations to the solved two-color fully-packed loop model 
\cite{kondev}. This predicts $C(x)\sim x^{-1/3}$ which is in perfect agreement with the data 
from our Monte Carlo simulation.

The different behavior of the defect-defect correlations leads to two different scenarios with 
respect to the separation of defects at small finite temperatures. A  separation of two defects 
to an infinite distance leads to an increase of the free energy, 
$\Delta F(\infty) \sim -T(\ln C(\infty)- \ln C(1))$. Consequentely, $\Delta F(\infty)$ is 
infinite for the defect on the square and honeycomb lattice (confinement) and remains finite 
for the triangular and diamond lattice (deconfinement).

The results are useful for the quantum version of the problem at the RK point of the Hamiltonian 
in general \cite{RK}. In particular, our results can be extended to the motion of strongly 
correlated spinless fermions on the checkerboard lattice \cite{frank}.
\acknowledgments
We are grateful to Peter Fulde, Kirill Shtengel and Dima Efremov for useful discussions and in particular to Roderich Moessner for many discussions and a careful reading of the manuscript.

\end{document}